\documentclass[journal]{IEEEtran}
\usepackage{cite}
\usepackage{amsmath, amssymb, amsfonts}
\usepackage{algorithm}
\usepackage[noend]{algorithmic}
\usepackage{graphicx}
\usepackage{textcomp}
\usepackage{xcolor}
\usepackage{color}
\usepackage{bm} 
\usepackage{subfigure} 
\usepackage{booktabs}
\usepackage{cleveref}
\usepackage{autobreak}
\usepackage{multirow}
\usepackage{booktabs}
\usepackage{authblk}
\usepackage{mathrsfs}
\usepackage{array}
\usepackage{makecell} 
\usepackage{float}
\usepackage{flushend}
\usepackage{soul}

\begin{document}

\title{Information Bottleneck-Inspired Type Based Multiple Access for Remote Estimation\\ in IoT Systems}

\author{
\IEEEauthorblockN{Meiyi Zhu, Chunyan Feng, \IEEEmembership{Senior Member, IEEE}, \\ \vspace{-1em} Caili Guo, \IEEEmembership{Senior Member, IEEE}, Nan Jiang, Osvaldo Simeone, \IEEEmembership{Fellow, IEEE}}
\vspace{-2em}
\thanks{This work was supported in part by the Fundamental Research Funds for the Central Universities (No. 2021XD-A01-1), in part by the National Natural Science Foundation of China (No. 92067202). The work of O. Simeone was supported by the European Research Council (ERC) under the European Union’s Horizon 2020 research and innovation programme, grant agreement No. 725731, by an Open Fellowship of the EPSRC with reference EP/W024101/1, and by the European Union through project CENTRIC (101096379). }
}
\maketitle

\begin{abstract}
Type-based multiple access (TBMA) is a semantics-aware multiple access protocol for remote inference. In TBMA, codewords are reused across transmitting sensors, with each codeword being assigned to a different observation value. Existing TBMA protocols are based on fixed shared codebooks and on conventional maximum-likelihood or Bayesian decoders, which require knowledge of the distributions of observations and channels. In this letter, we propose a novel design principle for TBMA based on the information bottleneck (IB). In the proposed IB-TBMA protocol, the shared codebook is jointly optimized with a decoder based on artificial neural networks (ANNs), so as to adapt to source, observations, and channel statistics based on data only. We also introduce the Compressed IB-TBMA (CIB-TBMA) protocol, which improves IB-TBMA by enabling a reduction in the number of codewords via an IB-inspired clustering phase. Numerical results demonstrate the importance of a joint design of codebook and neural decoder, and validate the benefits of codebook compression.
\end{abstract}


\vspace{-0.2cm}
\section{Introduction}
Conventional multiple-access channel (MAC) protocols for remote inference, which are central to many Internet-of-Things (IoT) deployments, are designed with the aim of recovering individual messages from the sensing devices. As a result, the required spectral resources grow proportionally to the number of devices. In the presence of a large number of devices, this may lead to an excessive communication overhead. This paper follows the work initiated by \cite{1_Liu_TBMA, 2_Tong_TBMA} to investigate MAC protocols for remote inference based on type-based multiple access (TBMA). By leveraging the semantic properties of the inference problem, TBMA requires spectral resources that grow with the diversity in observations, rather than with the number of devices.

In TBMA, all sensing devices share the same codebook, and each codeword is assigned to a single range of values for the sensors' observations. Therefore, devices making the same observation transmit the same codeword. Intuitively, assuming that the codewords are orthogonal, the receiver can estimate the number of devices making the same observations by measuring the power received for each codeword. This way, TBMA can be considered as a form of over-the-air computing (see, e.g., \cite{nazer2007computation}), whereby the receiver extracts information from multiple sufficient statistics obtained via the combination of signals transmitted by the devices. It can also be viewed as a form of joint source-channel coding, as it directly maps the observations of the devices to transmitted symbols in a way that directly optimizes the quality of the recovery of the ``semantics''  of the information source at the receiver \cite{DeepSC, farsad2018deep}.

\begin{figure}[t]
	\centering
	\includegraphics[width = 1\linewidth]{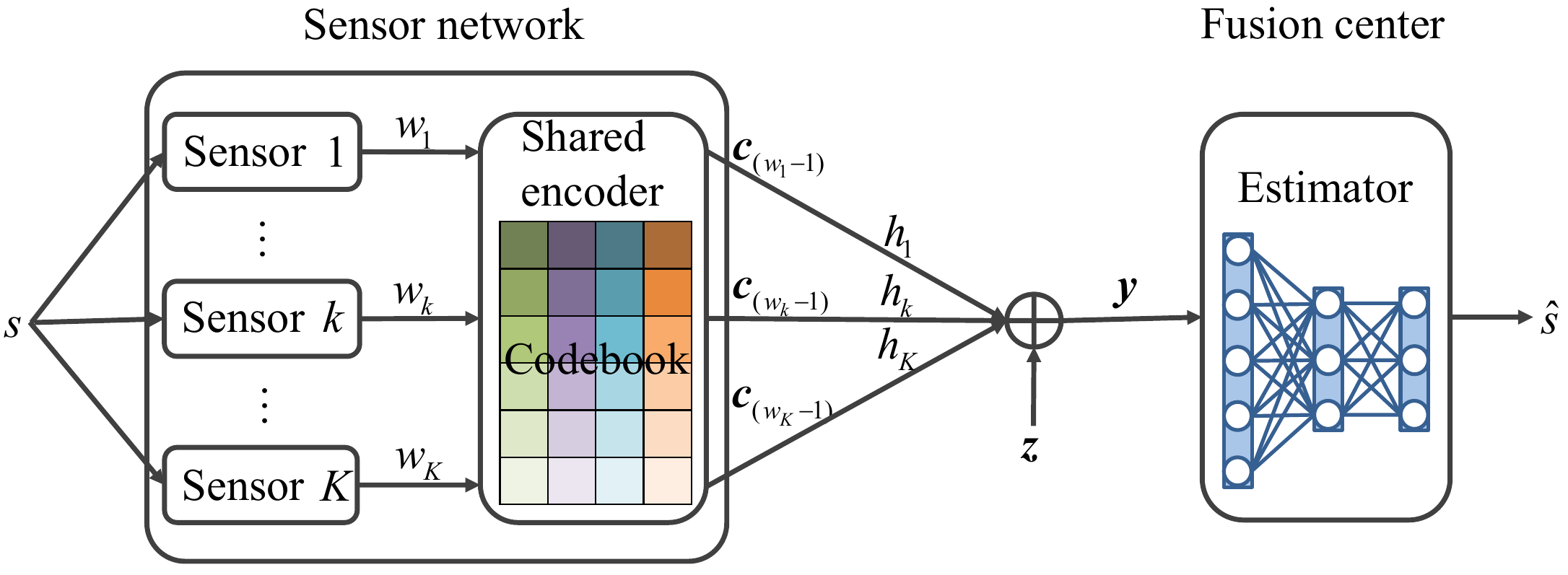}
	\caption{A TBMA-based system for remote estimation at a fusion center. We are interested in jointly optimizing the shared codebook used by the sensors for encoding, as well as the decoding operation at the fusion center.}
  \label{system model fig}
\end{figure}

The original papers \cite{1_Liu_TBMA, 2_Tong_TBMA} provided the theoretical underpinning for TBMA, and devised a practical variant of a maximum likelihood (ML) estimator at the fusion center. Reference \cite{3_multi_cell_FRAN} extended TBMA from a single-cell setting to multiple-cell systems, by leveraging in-cell TBMA in conjunction with inter-cell frequency reuse. TBMA was further extended to multi-event detection in random access scenarios, and approximate message passing (AMP) algorithms were designed on the assumption of sparse user activity in \cite{4_single_cell_codebook, 5_multi_cell_codebook}.

The described state of the art on TBMA has the following limitations, which we aim to address in this letter. First, existing works adopt fixed transmission codebooks, which are restricted to either orthogonal codewords -- requiring large transmission blocks -- or randomized constructions. Second, existing designs of TBMA detectors, such as ML and AMP, assume knowledge of the channel and possibly of the source statistics, which are practically rarely available.


To address these aspects, in this letter, we propose for the first time the use of a machine learning methodology for the end-to-end  design of TBMA protocols that are capable of operating under unknown source and channel statistics. Specifically, we adopt the information bottleneck (IB) principle as a design criterion, and introduce a novel IB-driven clustering approach to optimize the TBMA codewords. The proposed protocols are designed via a joint optimization of the shared codebook and an artificial neural network (ANN)-based decoder (see Fig. 1).

This way, this letter contributes to a growing line of work on the end-to-end design of communication systems based on joint source-channel coding principles and enabled by neural transceivers (see, e.g., \cite{JSCC, text_semantic, speech_semantic, multimodal_semantic}). The IB principle, initially formalized in the context of rate-distortion theory \cite{IB}, has found applications as a design criterion for communication systems in several settings, including cloud radio access networks and edge computing \cite{single_link_Dekorsy, C_RANs_Dekorsy, Shao_IB_communication_compression, IB_NOMA}. To the best of our knowledge, this is the first work that proposes the application of IB to the design of TBMA protocols.


\vspace{-0.2cm}
\section{System Model}\label{system_model}
As illustrated in Fig. \ref{system model fig}, we consider a wireless sensor network in which $K$ sensing devices individually observe a target random variable (RV) $s$, and communicate information about their local measurements to a fusion center (FC) using a joint source-channel coding strategy. As anticipated in the previous section, we adopt a TBMA protocol \cite{1_Liu_TBMA, 2_Tong_TBMA}, which directly maps observations to channel symbols, and we take a semantics-aware end-to-end design approach that jointly optimizes encoding and decoding at the FC, aiming at producing an estimate $\hat{s}$ of the quantity $s$.

\subsubsection{Encoding}
As in \cite{2_Tong_TBMA}, the target RV $s$ is discrete, and has an unknown probability distribution $p\left(s\right)$ on its support $\mathcal{S}$. Each sensor $k$ obtains a measurement $w_k$ that is related to the target RV $s$ through the unknown conditional distribution $p\left(w\mid s\right)$. Observations of different sensors are conditionally independent given $s$. Therefore, the vector of observations $\boldsymbol{w} = \left[w_1,\ldots , w_k ,\ldots , w_K \right]^{\mathrm{T}}$ is distributed as $\boldsymbol{w}\sim \prod_{k=1}^{K} p\left(w_k\mid s\right)$ when conditioned on $s$. The observations $w_k$ are also discrete, and take values, without loss of generality, in the set $\left\{0,1, \ldots, M-1 \right\}$.

According to the TBMA protocol, the sensors share a codebook for transmission to the FC that consists of $M$ codewords. Each codeword $\boldsymbol{c}_{m}\in \mathbb{C}^{N\times 1}$ is assigned to an observation value $m \in \left\{0, 1,\ldots , M-1 \right\}$ and consists of $N$ symbols. Accordingly, when observing $w_k=m$, device $k$ transmits codeword $\boldsymbol{c}_{m}$ across $N$ channel uses. Each codeword abides by power constraints $\| \boldsymbol{c}_m \| ^2\leq E$. In previous work \cite{2_Tong_TBMA, 3_multi_cell_FRAN}, codewords were chosen to be orthogonal, requiring $N\geq M$; while in \cite{4_single_cell_codebook, 5_multi_cell_codebook} all codewords were generated randomly with entries i.i.d. $\mathcal{C N} \left(0, E/N\right)$. In contrast, in this work, we leverage machine learning tools to adapt the codebook $\boldsymbol{C} = \left[\boldsymbol{c}_1, \ldots, \boldsymbol{c}_m, \ldots, \boldsymbol{c}_M \right] \in \mathbb{C} ^{N \times M}$ to the semantics of the problem defined by the unknown distributions $p(s)$, $p\left(w\mid s\right)$, as well as by the channel distribution and by the estimation loss function, both of which will be defined below.

\subsubsection{Channel}
The FC observes the coherent sum of the signals corrupted by noise and fading on the shared MAC. We denote $h_k$ as the flat-fading coefficient for sensor $k$. Furthermore, we introduce the one-hot vector $\boldsymbol{u}_k \in \mathbb{B}^{M\times 1}$, with all zero entries except for a single $1$ at the $\left(w_k+1\right)$-th position, for each observation $w_k$; as well as matrix $\boldsymbol{U} = [\boldsymbol{u}_1, \ldots, \boldsymbol{u}_k, \ldots, \boldsymbol{u}_K]\in \mathbb{C}^{M \times K}$ that gathers all observations. With these definitions, the received signal can be expressed as the superposition 
\begin{equation}\label{received_signal}
  \boldsymbol{y} =\sum_{k=1}^{K} h_{k} \boldsymbol{C} \boldsymbol{u}_{k}+\boldsymbol{z}
  =\boldsymbol{C} \boldsymbol{h}_{w}+\boldsymbol{z} \in \mathbb{C}^{N \times 1},
\end{equation}
where we have defined the vector $\boldsymbol{h}_{w} = \boldsymbol{U}\boldsymbol{h}$ with $\boldsymbol{h}=[h_1, \ldots, h_k, \ldots, h_K]^{\mathrm{T}} \in \mathbb{C}^{K\times 1}$, and $\boldsymbol{z} \in \mathbb{C}^{N\times 1}$ denotes the Gaussian noise for $N$-dimensional codewords with every entry drawn i.i.d. as $\mathcal{C N}\left(0, \sigma_z^2\right)$. Note that the effective channel vector $\boldsymbol{h}_{w}$ carries information about the observations made by the sensors, since each entry $m$ of the vector is non-zero only if at least one sensor $k$ has made the observation $w_k=m$. The channel vector $\boldsymbol{h}$ has an unknown distribution, which translates into an unknown conditional distribution $p(\boldsymbol{h}_w|\boldsymbol{w})$.

\subsubsection{Decoding}
Previous works\cite{2_Tong_TBMA,3_multi_cell_FRAN,4_single_cell_codebook,5_multi_cell_codebook} assumed that the receiver implements either ML or Bayesian estimators, both of which require knowledge of the distributions $p\left(w\mid s\right)$ and $p\left(\boldsymbol{y}\mid\boldsymbol{w}\right)$, with the latter also requiring $p(s)$. In contrast, in this work we introduce a data-driven ANN estimator $f_\theta\left(\cdot\right)$ parameterized by a vector $\theta$. The ANN $f_\theta\left(\cdot\right)$ takes as input the received signal $\boldsymbol{y}$, and produces as its output the conditional probability distribution $q\left(s\mid \boldsymbol{y}, \theta\right)$ over the values $\mathcal{S}$ of the target RV $s$. A hard estimate of the RV $s$ can hence be obtained as $\hat{s}=\arg \max_{s\in\mathcal{S}} q(s\mid\boldsymbol{y},\theta)$.

\vspace{-0.2cm}
\section{IB-TBMA Protocols} \label{IB_TBMA_protocals}
In this section, we introduce the proposed IB-based TBMA protocols, which implement an IB-inspired optimization of shared codebook and neural estimator at the FC. 

\vspace{-0.2cm}
\subsection{IB-based Optimization Problem}
The design goal of the TBMA system is to maximize the accuracy of the estimate $\hat{s}$ for the given number of available channel samples $N$. To characterize the trade-off between estimation accuracy and spectral efficiency via the number of channel uses, we adopt an IB-based design criterion. 

We write as $I\left(a;b\right) = \mathbb{E}_{p\left(a,b\right)}\left\{\log\left[p\left(a,b\right)/(p(a)p(b))\right]\right\}$ the mutual information (MI) between two random variables $a$ and $b$ with joint distribution $p(a,b)$ and marginal distributions $p(a)$ and $p(b)$ \cite{Inf_The}. With this definition, the proposed performance criterion is given
\begin{equation}\label{IB}
  \mathcal{L}\left(\boldsymbol{C}, \theta\right) = -I\left(\boldsymbol{y};s\right) + \beta I\left(\boldsymbol{y};\boldsymbol{w}\right),
\end{equation}
where $\beta>0$ is a Lagrange multiplier. In (\ref{IB}), the MI $I\left(\boldsymbol{y};s\right)$ gauges the amount of information available at the receiver about the target RV $s$. Therefore, maximizing this term directly affects the accuracy of the estimate $\hat{s}$ that can be extracted by the ANN from the received signal $\boldsymbol{y}$. In contrast, the MI $I\left(\boldsymbol{y};\boldsymbol{w}\right)$ captures the information that can be extracted from the received signal $\boldsymbol{y}$ about the individual observations of the sensors. Minimizing this term forces the codebook to contain more similar codewords. This, in turn, can be used, as discussed in Sec. \ref{compression} to compress the codebook $\boldsymbol{C}$, which can improve the spectral efficiency and robustness to noise of the remote estimation system.

\vspace{-0.2cm}
\subsection{Variational Optimization}
A direct minimization of the IB criterion (\ref{IB}) is infeasible since we assume no access to the distributions of the target RV $s$, of the observations $\boldsymbol{w}$ and of the channel vector $\boldsymbol{h}$. We hence adopt as performance criterion a variational upper bound on (\ref{IB}) that can be estimated based on samples from RV $s$, channels $\boldsymbol{h}$, and noise $\boldsymbol{z}$ \cite{VIB, simeone2022machine}.

\begin{algorithm}[t] 
  \caption{Codeword Clustering for CIB-TBMA}\label{codeword cluster}
  \hspace*{\algorithmicindent} \textbf{Input:} Codebook $\boldsymbol{C}$ and threshold $\gamma$\\
  \hspace{\fill}
  \hspace*{\algorithmicindent} \textbf{Output:} A set of clusters $\mathcal{C}$
  \begin{algorithmic}[1]
    \STATE {\textbf{Initialization:} Calculate distance matrix $\boldsymbol{D} = \left[d_{ij}\right] \in \mathbb{R}^{M \times M}$, where $d_{ij}=\|\boldsymbol{c}_i - \boldsymbol{c}_j\|_2$ denotes Euclidean distance between two codewords. Initialize the undirected graph $\mathcal{G} = (\mathcal{V},\mathcal{E})$ with $\mathcal{V} = \left\{v_1,\ldots,v_i,\ldots,v_M \right\}$, where vertex $v_i$ represents codeword $\boldsymbol{c}_i$, and the edge set $\mathcal{E} = \varnothing$ is initially empty. Initialize the set of clusters as the empty set, $\mathcal{C} = \varnothing$.}
    \FOR{every pair of nodes $v_i,v_j$ in $\mathcal{V}$}
      \IF{$e_{ij}\notin \mathcal{E}$ and $d_{ij} \leq \gamma$}
        \STATE{Add an edge $e_{ij}$ between nodes $v_i$ and $v_j$}
        \STATE{$\mathcal{E} = \mathcal{E} \cup \left\{e_{ij}\right\}$}
      \ENDIF
    \ENDFOR
    \REPEAT
    \STATE{Find the largest clique $\mathcal{G}^*=(\mathcal{V}^*, \mathcal{E}^*)$ as a subgraph of graph $\mathcal{G}$}
    \STATE{Add the clique vertex set as a new cluster to the cluster set, i.e., $\mathcal{C} = \mathcal{C} \cup \left\{\mathcal{V}^*\right\}$}
    \STATE{Remove $\mathcal{G}^*$, including all edges with one node in set $\mathcal{V}^*$ from the graph $\mathcal{G}$, obtain the smaller graph $\mathcal{G} = \mathcal{G}\backslash \mathcal{G}^*$}
    \UNTIL{$\mathcal{G} = \varnothing$}
  \end{algorithmic}
\end{algorithm}

To define the adopted performance criterion, we introduce the Kullback-Liebler (KL) divergence $D_{KL}\left(p\|q\right) = \mathbb{E}_{p(x)}\left[\log\left(p(x)/q(x)\right)\right]$ as a measure of the ``difference'' between distribution $p(x)$ and $q(x)$\cite{Inf_The}. Furthermore, we write  $p\left(\boldsymbol{y}\mid \boldsymbol{w}, \boldsymbol{C}\right) = \mathbb{E}_{p\left(\boldsymbol{h}_w \mid \boldsymbol{w}\right)}\left[p\left(\boldsymbol{y}\mid \boldsymbol{h}_w, \boldsymbol{C}\right)\right]$ for the distribution of the received signal (1) given the transmitted signal $\boldsymbol{w}$ and codebook $\boldsymbol{C}$, where the distribution $p\left(\boldsymbol{h}_w \mid \boldsymbol{w}\right)$ of the effective channel depends on the unknown channel distribution $p(\boldsymbol{h})$; $p\left(\boldsymbol{y} \mid s, \boldsymbol{C}\right)=\mathbb{E}_{p\left(\boldsymbol{w} \mid s\right) p\left(\boldsymbol{h}_w \mid \boldsymbol{w}\right)}\left[p\left(\boldsymbol{y} \mid \boldsymbol{h}_w, \boldsymbol{C}\right)\right]$ for the distribution of the received signal (1) given the target RV $s$ and the codebook $\boldsymbol{C}$; and $p\left(\boldsymbol{w}\right)=\mathbb{E}_{p\left(s\right)}\left[p\left(\boldsymbol{w} \mid s\right)\right]$ for the marginal distribution of the transmitted signals. 
 
We adopt as performance criterion the following upper bound $\mathcal{L}_V\left(\boldsymbol{C}, \theta\right)$ on the IB function (\ref{IB}), which is given by
\begin{equation}\label{loss}
  \mathcal{L}_V\left(\boldsymbol{C}, \theta\right) =D\left(\boldsymbol{C}, \theta\right)+\beta R\left(\boldsymbol{C}\right),
\end{equation}
with the ``distortion'' term, $D\left(\boldsymbol{C}, \theta\right)$, is a measure of estimation error, while the ``rate'' term, $R\left(\boldsymbol{C}\right)$, is an information-theoretic regularizer. Accordingly, the upper bound takes the form of a free energy metric \cite{simeone2022machine}, and the distortion and rate terms are derived using standard steps as \cite{VIB,simeone2022machine}  
\begin{align}\label{distortion}
    -I\left(\boldsymbol{y} ; s\right) 
    \leq -\mathbb{E}_{p\left(s\right) p\left(\boldsymbol{y} \mid s, \boldsymbol{C}\right)}\left[\log q\left(s \mid \boldsymbol{y}, \theta\right)\right] \triangleq D\left(\boldsymbol{C}, \theta\right)
\end{align} where the ANN-based estimator $q\left(s \mid \boldsymbol{y}, \theta\right)$ serves as a variational approximation to the distribution $p\left(s \mid \boldsymbol{y}\right)$; and
  \begin{align}\label{rate}
    I(\boldsymbol{y} ; \boldsymbol{w}) 
    &\leq\mathbb{E}_{p(\boldsymbol{w})}\left\{D_{K L}[p\left(\boldsymbol{y} \mid \boldsymbol{w}, \boldsymbol{C}\right) \| r(\boldsymbol{y})]\right\} \triangleq R(\boldsymbol{C}),
  \end{align}
where $r(\boldsymbol{y})$ is a fixed, arbitrary distribution on the space of the received signals.  

Accordingly, the distortion term (\ref{distortion}) measures the log-loss of the ANN estimator on average over the target RV, channels, and noise; while the rate term (\ref{rate}) is a regularizer that forces the distribution of the received signals not to be too dependent on the observation vector $\boldsymbol{w}$.

To evaluate the expectations in (\ref{distortion}) and (\ref{rate}), we apply Monte Carlo sampling and the reparameterization trick \cite{VAE} to attain an unbiased estimate and facilitate backpropagation as described in \cite{VIB} and \cite[Ch. 10]{simeone2022machine}.

\begin{figure}[t]
	\centering
	\includegraphics[width = 0.85\linewidth]{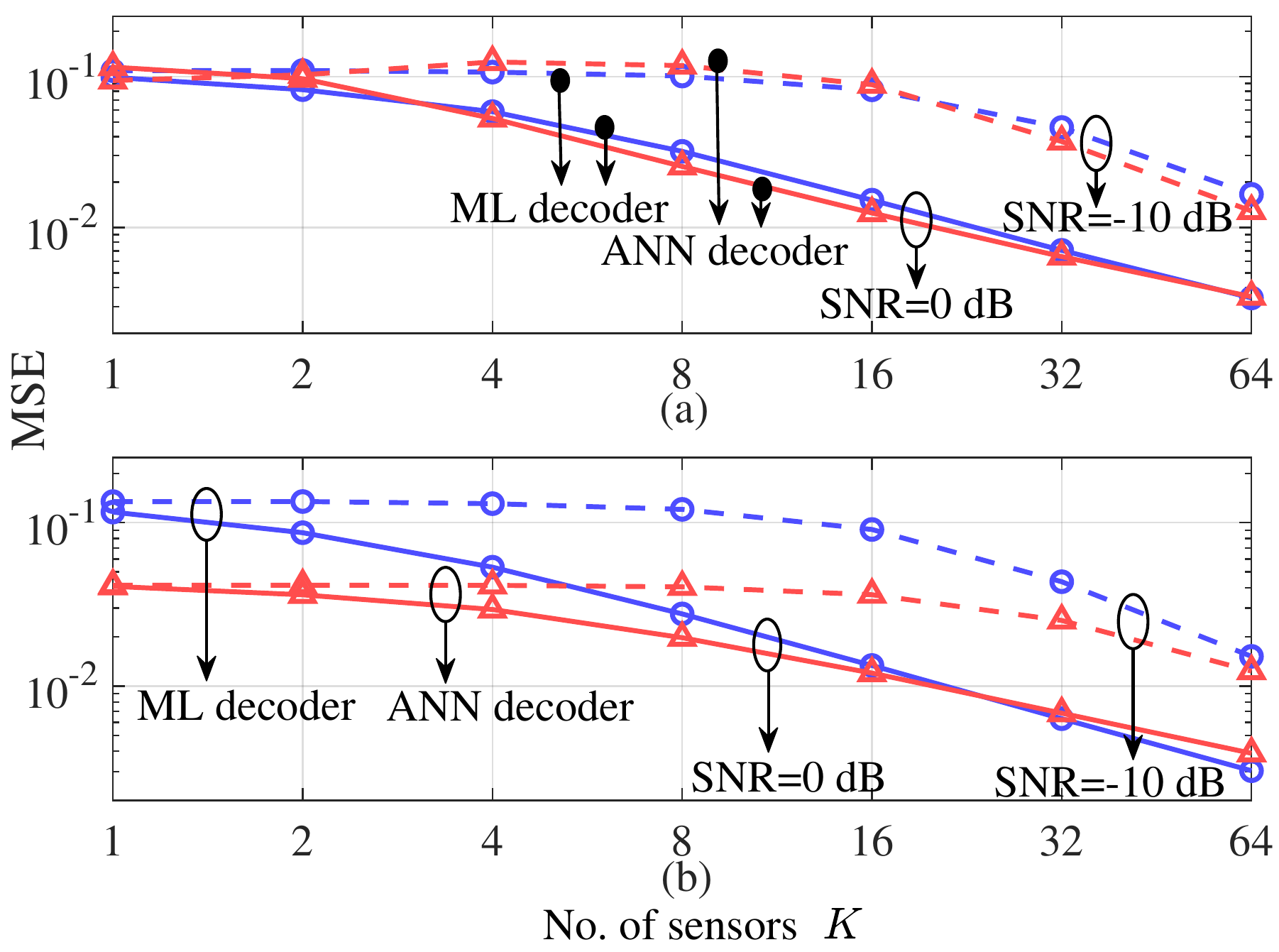}
	\caption{MSE of the approximate ML estimator introduced in \cite{2_Tong_TBMA} and of the proposed ANN-based decoder with fixed orthogonal codewords, within (a) uniform prior and (b) non-uniform prior under Gaussian channels.}
  \label{fixed codebook}
\end{figure}

\begin{figure*}[t]
	\centering
	\includegraphics[width = 0.95\linewidth]{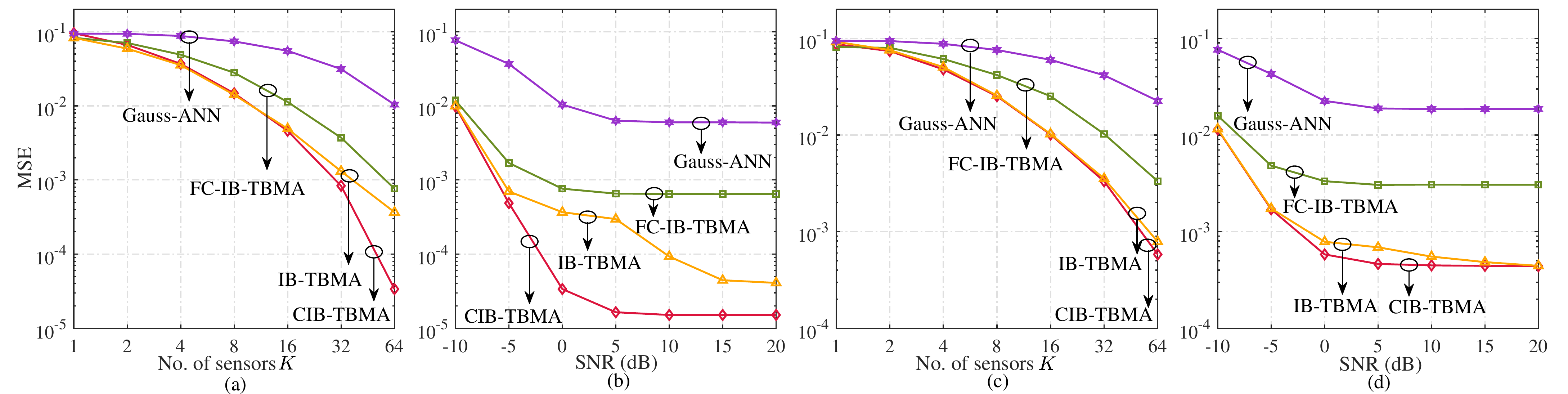}
  \caption{MSE of IB-TBMA, CIB-TBMA, as well as of the benchmark schemes Gauss-ANN and FC-IB-TBMA (see text for descriptions), under Gaussian channels (a), (b) and Rician fading channels (c), (d), versus the number of sensors with $\mathrm{SNR}=0$ $\mathrm{dB}$ in (a), (c) and versus the SNR for $K=64$ in (b), (d).}
  \label{random codebook}
  \vspace{-0.3cm}
\end{figure*}

\vspace{-0.2cm}
\subsection{IB-TBMA and CIB-TBMA}\label{compression}
We consider two implementations of the proposed IB-TBMA protocols: (\emph{i}) \emph{IB-TBMA}: In IB-TBMA, the minimization of (\ref{loss}) is carried out over codebook $\boldsymbol{C}$ and ANN decoder $\theta$ by setting $\beta=0$, hence targeting solely the accuracy criterion; and (\emph{ii}) \emph{Compressed IB-TBMA (CIB-TBMA)}: CIB-TBMA operates across two phases of minimization of the objective (\ref{loss}), with the goal of enhancing the detection performance of the ANN decoder. CIB-TBMA is described next.

As discussed in the previous section, the introduction of the rate term $R\left(\boldsymbol{C}\right)$ in the IB-based design objective (\ref{loss}) tends to reduce the dependence of the received signal $\boldsymbol{y}$ on the transmitted signals $\boldsymbol{w}$, yielding optimized codebooks with similar codewords. Leveraging this property of the IB-based criterion, CIB-TBMA aims to reduce the number of codewords. Reducing the number of codewords facilitates the detection task of the ANN at the receiver, particularly in the presence of a small number $N$ of channel uses. The key idea of the approach is to cluster the codewords by similarity after a first optimization of the variational criterion (\ref{loss}).

\subsubsection{Phase \uppercase\expandafter{\romannumeral1}: Uncompressed-codebook optimization} First, we address the problem of minimizing the IB criterion (\ref{loss}), with $\beta>0$, over the codebook $\boldsymbol{C}$ and model parameters $\theta$ of the ANN decoder $q\left(s\mid\boldsymbol{y},\theta\right)$. Then, we apply a clustering algorithm on the $M$ codewords, identifying the number $M'\leq M$ of clusters corresponding to groups of similar codewords. The specific proposed compression scheme is summarized in Algorithm \ref{codeword cluster}, and explained below at point 3).

\subsubsection{Phase \uppercase\expandafter{\romannumeral2}: Compressed-codebook optimization} In the second phase, we address again the problem of minimizing the IB criterion (\ref{loss}) over the codebook $\boldsymbol{C}$ and model parameters $\theta$ of the ANN decoder. However, this time we constrain the number of codewords to $M'$ by assigning the observations corresponding to one cluster to the same codeword. Furthermore, we set $\beta=0$ in order to focus on the minimization of the distortion term.

\subsubsection{Codeword clustering}
Inspired by \cite{graph_clustering}, we devise a graph-based clustering algorithm. The algorithm depends on a threshold $\gamma>0$ that is used to determine whether two codewords are ``similar'' or not. A larger threshold $\gamma$ tends to create fewer clusters, while a smaller $\gamma$ generally yields a larger number $M'$ of clusters. In the clustering algorithm, summarized in Algorithm \ref{codeword cluster}, a vertex is included in the graph for each codeword. Then, an edge is drawn between two vertices if the two corresponding codewords are ``similar'', where similarity is measured by comparing the Euclidean distance between the codewords to threshold $\gamma$. Finally, we repeatedly remove from the graph the largest clique, i.e., the largest subset of connected codewords, to form a new cluster. Ties can be broken arbitrarily.

\vspace{-0.2cm}
\section{Performance Evaluation and Conclusions} \label{results_and_conclusions}
In this section, we provide numerical examples in order to evaluate the performance of the proposed IB-TBMA  and CIB-TBMA protocols. We are specifically interested in quantifying the performance gains achievable via the use of a neural detector and via IB-based compression.

Throughout, we assume Gaussian channels with $h_k = 1$, as well as Rician fading channels $\boldsymbol{h} \sim \mathcal{C N}(\boldsymbol{\mu}, \sigma_h^2\mathbf{I})$ with an all-one mean $\boldsymbol{\mu}$ and variance $\sigma_h^2=1$. The average signal-to-noise ratio (SNR) is defined as $\mathrm{SNR} = E/\sigma^2_z$. The hyperparameter $\beta$ is chosen via cross-validation as $0.0009$. A single-layer network with tanh activation function scaled by $\sqrt{E/N}$ is used for the codebook to ensure the power constraints. The ANN decoder implements a two-layer perceptron with $16$ hidden neurons, ReLU activation function, and an output softmax layer with $|\mathcal{S}|$ neurons. Each output neuron reports the confidence level of the model in the target RV taking one of the $|\mathcal{S}|$ possible values. Minimization of the criterion (\ref{loss}) is carried out via the Adam optimizer with learning rate $0.001$ that decays tenfold every $10$ epochs during $100$ training epochs. The mean squared error (MSE) $\mathbb{E}[\left(\hat{s} - s\right)^2]$ is adopted as the performance metric, where $\mathbb{E}\left[\cdot\right]$ denotes the expectation over target RV $s$, channels $\boldsymbol{h}$ and noise $\boldsymbol{z}$.

\subsubsection{ANN Decoder vs. ML Decoder}
In this first experiment, we fix the shared codebook and evaluate the advantage of adopting a neural network decoder for TBMA as compared to the (approximate) ML decoder introduced in \cite{2_Tong_TBMA}. The  ANN decoder is trained by minimizing (\ref{loss}) with $\beta=0$. We follow the setting considered in \cite{2_Tong_TBMA} where we have $M=2$ and $N=2$; the target RV can take values in set $\mathcal{S} = \left\{0.1i \mid i = 1,2,\ldots,9\right\}$, and has either a uniform distribution $p(s) = 1/9$ or the non-uniform distribution $p(0.1) = p(0.9) = 0.05$, $p(0.2) = p(0.8) = 0.07$, $p(0.3) = p(0.7) = 0.12 $, $p(0.4) = p(0.6) = 0.16$, $p(0.5) = 0.2$; while the distribution of the observations, $p\left(w\mid s\right)$, is Bernoulli with probability $s$. The codebook is given by two arbitrary orthogonal codewords.

Fig. \ref{fixed codebook} evaluates the performance of the ANN decoder and of the approximate ML decoder by considering the uniform prior $p\left(s\right)$ (Fig. \ref{fixed codebook}(a)) and the non-uniform prior (Fig. \ref{fixed codebook}(b)) under a Gaussian channel. With a uniform prior, the ANN-based estimator performs very close to the ML estimator. In contrast, with a non-uniform prior, the ANN decoder outperforms the ML method, especially in the regime of fewer sensors. This is because the ANN estimator can incorporate the distribution of the target RV via training, which is particularly important in the regime of a limited number of observations.

\vspace{-0.1cm}
\subsubsection{IB-TBMA and CIB-TBMA}
We now evaluate the performance of IB-TBMA and CIB-TBMA, as introduced in the previous section. As benchmarks, we consider: (\emph{i}) \emph{Gaussian codebook with ANN decoder (Gauss-ANN)}: Only the decoder is optimized as discussed in the previous subsection, and the codebook uses fixed randomized Gaussian codewords as in \cite{4_single_cell_codebook,5_multi_cell_codebook}; and (\emph{ii}) \emph{IB-TBMA with fixed compression (FC-IB-TBMA)}: FC-IB-TBMA applies IB-TBMA on a modified system in which the $M$ observations are partitioned into $M'$ bins, with each bin containing adjacent values from the set $\{0,1,...,M-1\}$. The value $M'$ is chosen to be equal to that obtained by CIB-TBMA. 
The target RV takes values from the set $\mathcal{S} = \left\{0.2, 0.4, 0.6, 0.8\right\}$ with equal probability; and we set $N = 5$ and $M = 20$. The distribution $p\left(w\mid s\right)$ is such that $p\left(w\mid s\right)=1/20$ for even values of $w$, including $0$, for all $s\in \mathcal{S}$; while it is equal to a rescaled binomial distribution with parameters $(9,s)$ for odd values of $w$.

Fig. \ref{random codebook} plots the MSE of the considered methods versus the number of sensors, and versus the SNR for Gaussian channels and Rician fading channels. Both IB-TBMA and CIB-TBMA are seen to yield significant performance advantages over existing strategies with fixed codebooks, showing the importance of optimizing the shared codebook. Furthermore, codebook compression is found to be especially advantageous for a sufficiently large number of sensors and in the intermediate-SNR regime for Gaussian channels. In the latter regime, the performance bottleneck is set by channel noise, and codebook compression helps facilitate the detection of fewer codewords at the ANN. Compared with a fixed compression strategy FC-IB-TBMA, the gain of CIB-TBMA demonstrates the importance of an optimized clustering of the codewords that is applied jointly with codebook and decoder design.

\bibliographystyle{IEEEtran}
\bibliography{IB_TBMA}
\end{document}